\documentclass[sigplan,screen,nonacm,10pt]{acmart}
\renewcommand\footnotetextcopyrightpermission[1]{}
\usepackage{xspace}
\usepackage{makecell}
\usepackage{hyperref}
\usepackage{xr-hyper}
\usepackage{cleveref}
\crefname{section}{§}{§§}
\usepackage{algorithmicx}
\usepackage{algorithm}
\usepackage{algpseudocode}
\usepackage[utf8]{inputenc}
\usepackage{bbding}
\usepackage{pifont}
\usepackage{multicol}
\usepackage{multirow}
\usepackage{makecell}
\usepackage{subcaption}
\usepackage{graphicx}
\usepackage[toc,page]{appendix}
\usepackage{algorithm}
\usepackage{algpseudocode}
\newcommand{\Sys}[0]{psRL\xspace}

\AtBeginDocument{%
  }
 
\setcopyright{acmlicensed}
\copyrightyear{2018}
\acmYear{2018}
\acmDOI{XXXXXXX.XXXXXXX}
\acmConference[Conference acronym 'XX]{Make sure to enter the correct
  conference title from your rights confirmation emai}{June 03--05,
  2018}{Woodstock, NY}

\begin{document}

\title{\Sys: Efficient Training for Agentic AI via Training-Time Prefix Sharing
}

\author{
  \begin{tabular}{c}
    {\rm Mianjie Yu\textsuperscript{1},
    Zizhao Mo\textsuperscript{1}, Huanyu Qu\textsuperscript{1},
    Zhirong Qian\textsuperscript{1}, Huanle Xu\textsuperscript{1,\textdagger}}\\
    {\rm Cen Li\textsuperscript{2},
    Zifeng Zhao\textsuperscript{2}, Zhi Zhou\textsuperscript{2}, Jinhua Zhou\textsuperscript{2}, Jun Xie\textsuperscript{2},
    Chengzhong Xu\textsuperscript{1,\textdagger}}\\[0.25em]
    \textsuperscript{1}\textit{University of Macau} \quad
    % \textsuperscript{2}Sun Yat-sen University \quad
      \textsuperscript{2}\textit{Independent Researchers}
  \end{tabular}
} % end author


\begin{abstract}
In modern agentic AI training, the system bottleneck is shifting from rollout to update. Emerging sampling strategies such as tree-structured and step-wise RL greatly increase training sample volume while incurring relatively low marginal rollout cost, causing the update phase to dominate the end-to-end execution time. 
Crucially, this shift exposes a new optimization opportunity, as production traces reveal substantial prefix redundancy across training samples.

In this paper, we propose \Sys (\textbf{\underline{p}}refix \textbf{\underline{s}}haring for \textbf{\underline{RL}}), a new training system for agentic AI designed to exploit prefix redundancy among training samples. Leveraging the global visibility and data immutability inherent to the update phase, \Sys achieves efficient workload scheduling and memory management for distributed training. 
Specifically, \Sys introduces two novel prefix-sharing mechanisms that
enable flexible, fine-grained workload distribution
across GPU workers, simultaneously optimizing prefix reuse and achieving load balancing.
Moreover, \Sys implements a new underlying KV cache manager that facilitates adaptable block‑size allocation and dynamic KV caching, maximizing memory utilization while maintaining a high prefix hit rate. Evaluations using production traces demonstrate that \Sys outperforms existing systems by up to 5.2$\times$ in throughput. \textit{The source code will be publicly available soon}.

\end{abstract}

\maketitle
\begingroup
\renewcommand{\thefootnote}{\quad\quad\quad\textdagger}
\footnotetext{Corresponding authors: Huanle Xu (\texttt{huanlexu@um.edu.mo}) and Chengzhong Xu (\texttt{czxu@um.edu.mo}).}
\endgroup
\pagestyle{plain}

\vspace{-.5em}
\section{Introduction}
Agentic AI has fundamentally enhanced the reasoning capabilities of Large Language Models (LLMs), enabling them to tackle intricate tasks~\cite{guo2025deepseek,comanici2025gemini,llama4,claude4,wu2025agentic,zheng2025deepresearcher,prabhakar2025omniscience,zhou2025sweet}. Their efficiency stems primarily from advanced Reinforcement Learning (RL)~\cite{shao2024deepseekmath,yu2025dapo,schulman2017proximalpolicyoptimizationalgorithms,pmlr-v37-schulman15}.  A standard RL training pipeline comprises three distinct phases: \textit{rollout}, \textit{reward}, and \textit{update}. In the rollout phase, the model interacts with an environment to generate sequences, or trajectories, for a specific task. Subsequently, the reward phase evaluates action quality, assigning scalar scores to these trajectories. Finally, the update phase leverages the generated samples and corresponding reward signals to refine model parameters via gradient descent.

Under traditional linear, trajectory-wise sampling, autoregressive generation dominates end-to-end cost, positioning rollout latency as the primary optimization target. In modern RL training, however, the system bottleneck is shifting from rollout to update due to a rapidly changing workload landscape. Specifically, to improve sample diversity and mitigate reward sparsity, recent RL pipelines have increasingly adopted tree-structured sampling strategies~\cite{li2025treepo,hou2025treerl,ji2025tree,guo2025segment} and step-wise RL~\cite{feng2025group,wang2025stepsearch,luo2025agent,goldie2025synthetic,rllm2025}, where parallel exploration is launched from shared decision prefixes. 
While these approaches significantly increase the number of training samples derived from a single trajectory and often enhance convergence quality—for instance, in industrial Digital Twin Network (DTN) operations agent, step-wise RL achieves nearly \textbf{2$\times$} higher convergence performance than linear trajectory-wise RL—they expand the solution space and training sample volume at low marginal rollout cost, yet introduce a severe and often overlooked system overhead. Our analysis of production workloads and internal benchmarks reveals a fundamental shift in the bottleneck: update overhead has grown by more than \textbf{5}$\times$ and now accounts for over \textbf{50}\% of end-to-end RL training time, dominating overall system latency.

Fortunately, this algorithmic shift conceals a significant optimization opportunity: generated training samples exhibit high prefix redundancy. This phenomenon is intrinsic to the sampling methodologies themselves. In tree-structured RL, multiple parallel branches diverge from a common parent node, resulting in distinct samples sharing identical extensive prefixes.  Similarly, step-wise RL creates a recursive data dependency, where a sample at step $t$ essentially encompasses the token sequence of step $t$-1. Our analysis of both open-source and industrial production traces reveals that prefix matching rates frequently exceed \textbf{90}\% in agentic workloads. This indicates that current training systems~\cite{narayanan2021efficient,zhao2023pytorch,rasley2020deepspeed} treating samples as isolated sequences incur massive waste by repeatedly calculating and storing identical data segments.

Drawing inspiration from the success of prefix sharing in inference systems~\cite{kwon2023efficient,zheng2024sglang,ye2024chunkattention,li2025hotprefix}, we propose to adapt and extend these primitives to the training domain. 
However, realizing this optimization within existing infrastructures presents two primary challenges. First, efficient workload scheduling requires increasing the prefix match rate while simultaneously achieving load balancing—a fundamentally difficult combination. Disparities in workload across micro-batches, driven by varying sequence lengths and the extent of prefix sharing, introduce severe pipeline bubbles. Second, traditional memory management poses a critical bottleneck. Allocating KV states with fixed-size blocks fundamentally mismatches the highly variable lengths of reused prefixes, forcing a strict trade-off between memory access overhead and memory consumption overhead. Moreover, existing caching policies struggle to balance minimizing the KV state footprint with maintaining a high prefix hit rate.

In this paper, we introduce \Sys to enable effective prefix sharing and address the challenges in agentic RL training. The architecture of \Sys leverages a fundamental characteristic of training: the training phase exhibits two key properties—\textit{Global Visibility} and \textit{Data Immutability}. Since the entire input dataset is known upfront and its distribution remains static within each iteration, \Sys unlocks a broad design space for optimization across the system stack.

To tackle the challenges of workload scheduling, \Sys proposes two novel prefix-sharing mechanisms:
% enabling flexible, fine-grained workload distribution: 
inter-batch and self-sequence sharing. 
Inter-batch sharing allows sequences with shared prefixes to be distributed across different micro-batches while reusing the computed KV states, thereby providing high scheduling flexibility. Self-sequence sharing partitions a single long sequence into multiple contiguous chunks assigned to different micro-batches. By ensuring that suffix tokens reuse the KV states of prefix tokens, this mechanism facilitates fine-grained distribution of the computational load. Building on these two mechanisms, \Sys implements a hierarchical workload distribution strategy.  Globally, \Sys constructs semantic groups consolidating sequences with identical prefixes, employing efficient algorithms to partition them into evenly distributed workloads across training workers. Within each worker, \Sys utilizes token-wise micro-batching to slice sequences at the token level, significantly reducing pipeline bubbles caused by internal workload imbalances.

\Sys also designs an adaptive KV cache manager with two new mechanisms that effectively overcome the bottlenecks of traditional memory management. Specifically, the Adaptive Block Allocation mechanism allocates variable-sized contiguous memory blocks via analyzing the precise lengths of shared prefixes, thereby eradicating internal fragmentation~\cite{kwon2023efficient,zheng2024sglang} and improving memory access efficiency. Furthermore, to reduce the KV cache memory footprint while maintaining a high prefix sharing rate, the Dynamic Block Caching mechanism employs a strict Just-in-Time policy. By analyzing the prefix tree constructed during rollout, this policy proactively caches KV blocks required for future operations and promptly evicts unreferenced blocks, maximizing system throughput and computational efficiency.

We implement \Sys on top of veRL~\cite{sheng2025hybridflow}, with the primary modifications made to Megatron-LM~\cite{narayanan2021efficient}, the underlying training engine used in the RL update phase. Notably, we extend the scope of our prefix sharing and optimization to accelerate the \textit{reference model} forward passes and the computation of \textit{old policy log-probabilities}, both of which are critical in RL pipelines.
We also conduct evaluations using real-world production traces from industry agentic workloads, comparing \Sys against state-of-the-art training systems. Experimental results demonstrate that \Sys achieves up to 5.2$\times$ improvement in training throughput. To summarize, we have made the following contributions in this paper:
\vspace{-.3em}
% In summary, this paper makes the following contributions: 
\begin{itemize}
\item We identify the fundamental shift in Agentic AI training towards update-bound workloads and quantitatively demonstrate the high prefix redundancy inherent in tree-structured and step-wise RL. 
\item We design two novel prefix-sharing mechanisms that enable flexible, fine-grained workload distribution across GPU workers, simultaneously optimizing prefix reuse and achieving load balancing.
\item We implement a new underlying KV cache manager that facilitates adaptable block‑size allocation and dynamic KV caching, maximizing memory utilization while maintaining a high prefix hit rate.
 
\end{itemize}

\vspace{-.7em}
\section{Background and Motivation}

\subsection{Agentic Training Pipelines}
The training of Agentic AI often operates as an iterative RL process consisting of multiple repeated cycles~\cite{shao2024deepseekmath,yu2025dapo,schulman2017proximalpolicyoptimizationalgorithms,pmlr-v37-schulman15,zheng2025groupsequencepolicyoptimization,zheng2025prosperity}.  Each cycle typically comprises three distinct stages:

\noindent\textbf{\textit{Rollout:}} The LLM functions as a policy network, generating responses or action trajectories from input prompts. 
% From a system perspective, 
This phase performs standard batched LLM inference, producing token sequences via autoregressive decoding.

\noindent\textbf{\textit{Reward:}} The generated responses are evaluated to assign scalar scores. This scoring is typically performed using either rule-based functions or learned reward models that approximate human preferences.

\noindent\textbf{\textit{Update:}} Leveraging the generated data and rewards, the system calculates the loss and performs backpropagation. This involves computing parameter gradients and applying optimizer steps to update weights of LLMs. 
% refining the policy for the next cycle.
\vspace{-.3em}
\subsection{The Rollout-Centric Optimization Paradigm}
Following the standard training pipelines introduced above, prior system optimizations in RLHF and agent training~\cite{tan2026orchestrrl,qin2025seer,wu2025rlboost,team2025kimi,qu2025copris,zhong2025streamrl,chen2025respec,cheng2025fast,he2025history,fu2025areal} overwhelmingly target the rollout phase. This focus stems from a fundamental characteristic of traditional RL workloads: they predominantly rely on short, independent, and linear trajectories. Under these regimes, the sequential, memory-bandwidth-bound nature of autoregressive token generation emerges as the primary bottleneck dictating end-to-end training latency. To alleviate generation overhead, Kimi-K2~\cite{team2025kimi} employs aggressive trajectory truncation, while StreamRL~\cite{zhong2025streamrl} tackles load imbalance via dynamic batch resizing. CoPRIS~\cite{qu2025copris} bounds these inefficiencies by terminating generation once sufficient samples are gathered and recycling unfinished trajectories for subsequent iterations. Building on this speculative decoding (SD)~\cite{leviathan2023fast}, SpecActor~\cite{cheng2025fast} deploys a lightweight draft path to rapidly propose tokens, ensuring algorithmic fidelity through parallel verification against the target model. ReSpec~\cite{chen2025respec} dynamically tunes SD configurations and continuously evolves the drafter via knowledge distillation. Furthermore, systems like RhymeRL~\cite{he2025history} and Seer~\cite{qin2025seer} exploit similarity of samples to maximize rollout throughput by SD.

\noindent\textbf{The Architectural Blind Spot.} 
While these techniques successfully optimize the generation step, they are intrinsically predicated on the assumption of isolated and linear sample trajectories. However, the emergence of modern agentic workloads fundamentally shatters this assumption, rendering these rollout-centric optimizations insufficient.

\begin{figure}
    \centering
    \includegraphics[width=1\linewidth]{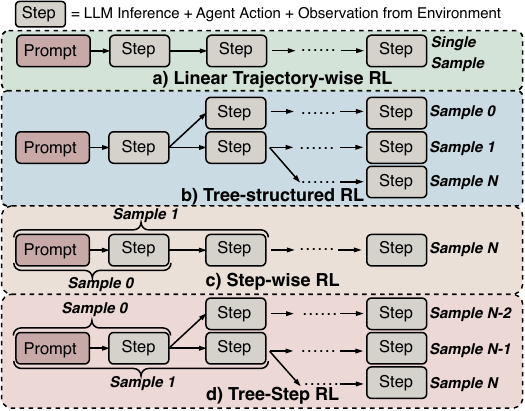}
    \vspace{-2em}
    \caption{Evolution of data sampling in agentic training.  a) Linear Trajectory-wise RL treats full interaction sequences as single atomic samples. To improve sample efficiency and reward density, modern workloads shift to structured paradigms: b) Tree-structured RL explores parallel branches, c) Step-wise RL decomposes trajectories into granular, cumulative samples, and d) Tree-Step RL combines both. }
\vspace{-1.em}
    \label{fig:Multi-step}
\end{figure}

\begin{figure*}[ht]
    \centering
    % --- (a) Reward Comparison ---
    \begin{subfigure}{0.214\textwidth}
        \centering
        \vspace{-.5em}
        \includegraphics[width=\linewidth]{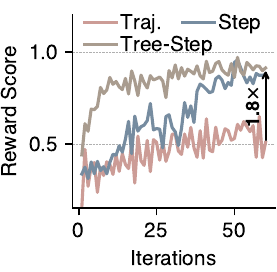}
        \vspace{-1.5em}
        \caption{Reward convergence.}
        \label{fig:reward_comparison}
    \end{subfigure}
    % \hfill % 自动填充间距
    % --- (b) RL Overhead ---
    \begin{subfigure}{0.756\textwidth}
        \centering
        \includegraphics[width=\linewidth]{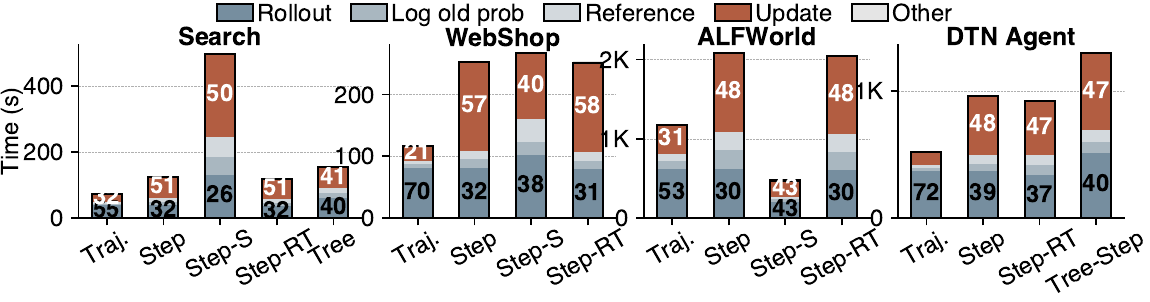}
        \vspace{-1.5em}
        \caption{Breakdown of end-to-end RL time across diverse agent workloads (percentages shown inside the bars).}
        \label{fig:RL_overhead}
    \end{subfigure}

\vspace{-1em}
    \caption{The algorithmic benefits and system bottlenecks of structured agentic sampling. a) Unlike traditional Trajectory-wise RL, Step-wise sampling resolves credit assignment issues, achieving significantly higher final rewards. b) However, shifting to structured topologies fundamentally alters the system workload,  inflating Update phase latency due to massive samples.
    % \vspace{-.5em}
    }
    \label{fig:ABC}
\end{figure*}

\begin{table*}[htbp]
    \centering
    % \vspace{-.5em}
    \caption{Prefix Match Rate (PMR) of sample data across diverse agent workloads}
    \label{tab:prefix_match_rate}
    \vspace{-.5em}
    % 如果表格太宽，使用 resizebox 将其缩放到文本宽度
    \resizebox{\textwidth}{!}{%
    \begin{tabular}{lccccccccccccccccc}
        \toprule
        \textbf{Agent} & \multicolumn{5}{c}{\textbf{Search}} & \multicolumn{4}{c}{\textbf{WebShop}} & \multicolumn{4}{c}{\textbf{ALFWorld}} & \multicolumn{4}{c}{\textbf{DTN Agent}} \\
        \cmidrule(lr){2-6} \cmidrule(lr){7-10} \cmidrule(lr){11-14} \cmidrule(lr){15-18}
        
        % 第二行：Mode
        \textbf{Mode} & Traj. & Step & Step-S & Step-RT & Tree & Traj. & Step & Step-S & Step-RT & Traj. & Step & Step-S & Step-RT & Traj. & Step & Step-RT & Tree-Step \\
        \midrule
        
        % 第三行：Prefix Match Rate
        \textbf{PMR (\%)} & 51.85 & 72.56 & 25.44 & 68.44 & 53.81 & 13.86 & 90.22 & 21.79 & 88.85 & 11.97 & 94.51 & 42.63 & 94.44 & 53.01 & 88.92 & 89.13 & 62.24 \\
        \bottomrule
    \end{tabular}%
    }
    % \vspace{-1em}
\end{table*}

\subsection{Status Quo: The Shift to Structured RL Sampling}

The pursuit of higher algorithmic accuracy and sample efficiency has driven the adoption of complex, structured sampling strategies in industry. As illustrated in Fig.~\ref{fig:Multi-step}, we classify agent training workloads into four distinct patterns.

\noindent\textbf{a) Linear trajectory-wise RL.} This is the foundational paradigm used in standard RLHF. The algorithm treats the entire interaction trajectory—comprising multiple steps of thoughts, actions, and observations—as a single, atomic training sample. However, this approach is plagued by the sparse reward problem and high sampling overhead~\cite{hou2025treerl,ji2025tree,li2025treepo}, as each sample requires a full, independent rollout generation.

\noindent\textbf{b) Tree-structured RL.} To address sample inefficiency, Tree-structured sampling (e.g., ByteDance's TreePO~\cite{li2025treepo}, TreeRL~\cite{hou2025treerl}) introduces parallel branching at critical decision points. In this paradigm, each complete path from the root to a leaf node constitutes a distinct training sample. By exploring multiple branches from a shared parent node, this mode leverages prefix reuse to amortize the rollout cost.

\noindent\textbf{c) Step-wise RL.} To tackle reward sparsity, Step-wise methods (e.g., Microsoft's LightningRL~\cite{luo2025agent}, DeepMind's SWiRL~\cite{goldie2025synthetic}, SenseTime's StepSearch~\cite{wang2025stepsearch}, GiGPO~\cite{feng2025group},  rLLM~\cite{rllm2025}) decompose a long trajectory into granular step-wise samples. Each step is scored and treated as an independent training unit, providing dense supervision signals. Consequently, every partial trajectory—extending from the root to any intermediate node—serves as an independent training sample. To further optimize context length and computational cost, several variants exist: 1)
 \textit{Step w/ Remove Think} (e.g., MiniWoB, FrozenLake): Removes internal "thought" chains from the history before feeding into the next step.
 2) \textit{Step w/ Summary} (e.g., Search~\cite{jin2025search}, ALFWorld~\cite{shridhar2020alfworld}): Compresses the historical context into a summary before feeding into the next step.

\noindent\textbf{d) Tree-Step RL.} Combining the benefits of parallel exploration and dense supervision, Tree-Step methods (e.g., SPO~\cite{guo2025segment}, Alibaba's Tree-GRPO~\cite{ji2025tree}) represent the state-of-the-art agentic training. This mode performs parallel branching at each step while simultaneously assigning rewards to individual actions. {As a result, every node in the generated search tree (spanning from the root to any leaf or intermediate node) constitutes an independent training sample.}

\noindent\textbf{The Algorithmic Superiority of Structured Sampling.} 
The industry-wide transition to structured sampling is fundamentally driven by its superior model performance in complex tasks. As demonstrated in Fig.~\ref{fig:reward_comparison}, when training an industrial DTN agent built upon Qwen3-235~\cite{qwen2025qwen25technicalreport} (detailed in Sec.~\ref{sec:Models and configuration}), the \textit{Step-wise} and \textit{Tree-Step} paradigm~\cite{feng2025group,ji2025tree} drastically outperforms the traditional \textit{Trajectory-wise} baseline~\cite{shao2024deepseekmath}. Because traditional RL defers the reward signal to the end of a long interaction sequence, it suffers from severe credit assignment issues; consequently, its reward fluctuates heavily and stagnates around 0.5. In contrast, by explicitly decomposing the trajectory and providing dense supervision, the Step-wise and Tree-Step modes enable highly stable convergence, achieving a nearly $1.8\times$ higher final reward within 60 iterations.

\vspace{-.5em}
\subsection{The Bottleneck Shift: From Rollout-Bound to Update-Bound}
While unlocking substantial algorithmic gains, the structured sampling strategies introduce a severe and largely overlooked cost on the training pipeline. In particular, the exponential increase in generated training samples fundamentally reshapes the performance bottleneck of agent training. To quantify this impact, we conduct a comprehensive experiment across diverse production-grade agents, including Search~\cite{jin2025search}, 
WebShop~\cite{yao2022webshop}, ALFWorld~\cite{shridhar2020alfworld}, and a large-scale industrial DTN Agent. We evaluate multiple representative sampling strategies: linear trajectory-wise RL (Trajectory)~\cite{sheng2025hybridflow}, tree-structured RL (Tree)~\cite{hou2025treerl}, step-wise RL (Step)~\cite{feng2025group}, and its variants Step with Removed Thinking (Step-RT), Step with Summarization (Step-S), and Tree-Step~\cite{ji2025tree}. Detailed experimental settings are summarized in Tab.~\ref{tab:configuration} and Sec.~\cref{sec:Experiment Setup}.
Fig.~\ref{fig:RL_overhead} illustrates the breakdown of end-to-end RL time across these workloads. First, except for agents utilizing summarization (Step-S), employing step-wise or tree-structured RL significantly inflates end-to-end RL latency compared to linear trajectory baselines. For instance, in DTN Agent, the total RL time for structured sampling strategies is 1.78$\times$ to 2.5$\times$ that of the linear trajectory baseline.

Crucially, this latency surge is not caused by slower rollout execution. 
Instead, the overhead stems entirely from the Update phase. Once step-wise or tree-structured RL is adopted, update time increases disproportionately. For example, in the WebShop workload, the update phase for step-wise RL increases by over 5$\times$ compared to linear trajectory sampling, while the rollout time remains virtually unchanged.
Consequently, a consistent bottleneck shift emerges across all agents. Under traditional trajectory-wise training, rollout clearly dominates RL time, accounting for 53\%--72\% of the total runtime. However, in structured sampling regimes, the update phase rapidly becomes the dominant cost, contributing up to 40\%--58\% of the total RL time. This contrasts starkly with the linear trajectory baseline, where the update phase typically consumes only 19\%--32\%. This shift indicates that agent training is no longer bottlenecked by rollout generation but by model updates, exposing a fundamental inefficiency in how current systems handle the training phase.

\subsection{Opportunity: Exploiting Prefix Redundancy}

While the shift to update-bound workloads presents a performance bottleneck, it simultaneously exposes a massive optimization opportunity.
Unlike standard pre-training data which is typically unique, agentic training samples generated via structured sampling share extensive common prefix.

To assess the magnitude of this potential, we analyzed the \textit{Prefix Match Rate}—defined as the percentage of prefix tokens in a batch that are identical to those in other samples.
Tab.~\ref{tab:prefix_match_rate} presents the results across different agents and sampling strategies.
In the traditional \textit{Trajectory} mode, redundancy is relatively low (e.g., $\sim$12\% for ALFWorld and $\sim$14\% for WebShop), as each sample is a distinct, full-length rollout.
 However, structured sampling fundamentally alters this profile. In \textit{Step} mode, where a trajectory is sliced into cumulative segments, the prefix match rate skyrockets to \textbf{94.51\%} in ALFWorld and \textbf{90.22\%} in WebShop.
 Even in high-complexity industrial scenarios like the \textit{DTN Agent}, the Step-wise redundancy remains exceptionally high at \textbf{88.92\%}, and Tree-Step maintain a robust match rate of \textbf{62.24\%}.

\begin{figure}
    \centering
\includegraphics[width=1\linewidth]{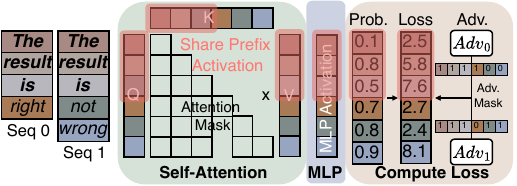}
\vspace{-2em}
    \caption{Training-Time Prefix Sharing.}
    \vspace{-1em}
    \label{fig:Shared_KV_base}
\end{figure}

\noindent\textbf{Prefix sharing in training.} Motivated by the high prefix redundancy in agentic training samples, we propose \textit{Training-Time Prefix Sharing} to accelerate the update stage. Since training requires rigorous computation of loss functions and gradient backpropagation, we leverage two key mechanisms to ensure correctness (as shown in Fig.~\ref{fig:Shared_KV_base}):
1) Attention Masking: Applies a sequence‑specific attention mask that strictly follows the original autoregressive dependency of each sequence. This guarantees that, although QKV are shared, the attention results remain identical to those obtained from independent computation. 2) Advantage Masking: Introduces an advantage mask during loss computation, as different sequences may have distinct advantage values. This mask ensures that each token only contributes to the loss corresponding to its own sequence, preventing cross‑sequence interference and preserving correct gradient flow.

% \vspace{-.5em}
\section{System Overview}

\subsection{Design Challenges} 
\label{challenges}
Leveraging prefix sharing can significantly reduce redundant computations and accelerate end-to-end training throughput. However, realizing this optimization within existing infrastructures presents two key challenges.

\textbullet\ \textit{The workload scheduling of training samples poses an immediate challenge.} Simultaneously increasing the prefix match rate and achieving load balancing is inherently difficult. At the global level, the system must orchestrate data distribution across all training workers to achieve both a high internal prefix match rate and a balanced workload among them. At the local level, the micro-batching strategy within each worker further complicates workload management. Specifically, both the extent of prefix sharing and the sequence lengths within a single micro-batch dictate its execution time. Consequently, workload disparities among different micro-batches introduce pipeline bubbles, which subsequently prolong the overall execution time of the worker. 
% As a result, optimizing prefix reuse and achieving load balancing emerge as tightly coupled problems.

\textbullet\ \textit{Memory management introduces another critical bottleneck.} Current systems typically allocate KV state using fixed-size blocks—a design fundamentally misaligned with the highly variable lengths of reused prefixes. This mismatch forces a hard trade-off: excessively small block sizes increase memory access overhead, while overly large blocks cause redundant prefix storage and introduce internal fragmentation. Furthermore, the system must balance the trade-off between reducing the KV state footprint and maintaining a high prefix hit rate. Storing an excessive number of KV states constrains available memory for batching, thereby reducing throughput. Conversely, aggressively evicting KV states diminishes prefix utilization, which impairs computational efficiency.

\subsection{System Architecture}
To support effective prefix sharing while addressing its inherent challenges in agentic RL training, we introduce \Sys. The two key properties inherent to the training phase—\textit{Global Visibility} and \textit{Data Immutability}—unlock significant opportunities for deterministic optimization in workload scheduling and memory management.

\noindent\textbf{Workload scheduling.} 
To tackle the challenges of workload distribution, \Sys proposes two novel prefix-sharing mechanisms that enable flexible and fine-grained workload distribution: inter-batch sharing and self-sequence sharing. These two sharing mechanisms enable high scheduling flexibility that facilitate fine-grained distribution of the computational load. 
Building on these mechanisms, \Sys implements a hierarchical workload distribution strategy. At the global level, \Sys constructs semantic groups to consolidate sequences sharing identical prefixes across the global batch, and employs efficient algorithms to partition these groups into evenly distributed workloads across workers. At the local worker level, it employs token-wise micro-batching, which slices sequences into micro-batches at the token level, significantly reducing pipeline bubbles caused by workload imbalances within each worker. 

\begin{figure}
    \centering
    \includegraphics[width=1\linewidth]{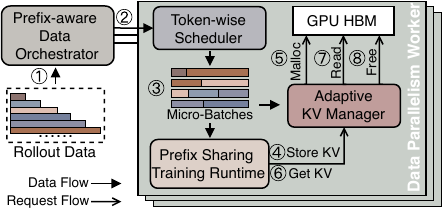}
    \vspace{-1em}
    \caption{System architecture of \Sys.}
     \vspace{-1.em}
    \label{fig:Overview}
\end{figure}

\begin{figure*}[t]
    \centering
\includegraphics[width=0.99\linewidth]{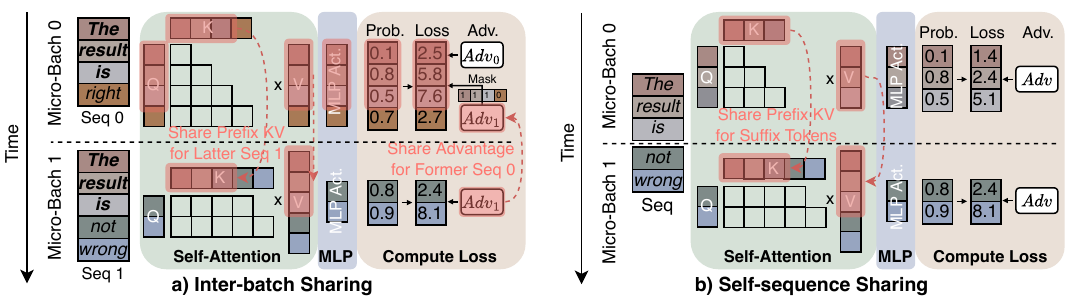}
\vspace{-1em}
    \caption{Mechanisms for Training-Time Prefix Sharing. a) \textit{Inter-batch Sharing} caches prefix KV states and  advantage values across sequential micro-batches, eliminating redundant computations. b) \textit{Self-sequence Sharing} partitions a long sequences across multiple micro-batches, reusing cached KV to bound peak memory footprint and eliminate pipeline bubbles.}
    % \vspace{-.8em}
    \label{fig:Shared_KV}
\end{figure*}

\noindent\textbf{Memory management.} Leveraging global visibility and data immutability, \Sys introduces another two mechanisms to overcome the bottlenecks of traditional memory management. First, to eliminate the rigid trade-offs imposed by fixed-size paging, the Adaptive Block Allocation mechanism allocates variable-sized contiguous memory blocks based on the precise lengths of shared prefixes. 
% Since this alignment ensures blocks correspond exactly to prefix boundaries, it simultaneously eliminates internal fragmentation and preserves high memory access efficiency without redundant storage. 
Second, the Dynamic Block Caching mechanism replaces conventional LRU-based caching with a strict Just-in-Time policy. By analyzing the prefix tree constructed from
rollout data, this mechanism proactively caches KV blocks required by future training samples and eagerly evicts blocks that are no longer referenced, thus eliminating wasted GPU memory.

\noindent\textbf{System workflow.}
As shown in Fig.~\ref{fig:Overview}, the system workflow proceeds as follows:
\textcircled{1} The {Orchestrator} ingests raw rollout data and constructs a global data partition to identify prefix reuse opportunities.
\textcircled{2} It dispatches optimally partitioned data subsets to each training worker.
\textcircled{3} Inside each worker, the Planner analyzes the assigned data, slicing data into optimized micro-batches and generating a deterministic execution schedule for the Runtime.
\textcircled{4}  During execution, the Runtime sends computed KV states to the Adaptive KV Cache Manager.
\textcircled{5}  The Manager proactively preserves the KV states required by subsequent micro-batches within HBM.
\textcircled{6}  When a subsequent micro-batch requires shared KV, the Runtime requests the KV blocks.
\textcircled{7}  The Manager retrieves the data directly from the high-speed cache.
\textcircled{8}  Upon the completion of the last dependent sequence, the Manager automatically frees the memory to reclaim space.

% \vspace{-.3em}
\section{Workload Scheduling}
This section details \Sys's hierarchical workload distribution scheme. We first introduce two key prefix-sharing mechanisms that enable fine-grained, flexible workload management. Building on these primitives, we  describe the global scheduling algorithm and present a token-wise micro-batching strategy used at the local worker level.

% \vspace{-.2em}
\subsection{Flexible Prefix Sharing Mechanisms}

\subsubsection{Inter-batch Sharing.}\label{sec:Inter-batch Sharing}
Inter-batch sharing addresses prefix redundancy across micro-batches, permitting shared prefixes to be reused within the same training worker. This mechanism enhances the prefix reuse ratio while simultaneously supporting flexible workload distribution. Specifically, as shown in Fig.~\ref{fig:Shared_KV}(a), when sequences sharing an identical prefix are assigned to different micro-batches, the system ensures that the earliest micro-batch computes and caches the KV state of the prefix. Subsequently, later micro-batches reuse this stored KV state and the associated attention results, thereby completely bypassing repeated computations of the prefix. Crucially, the earlier micro-batch also receives the advantage values of subsequent sequences and computes the corresponding loss utilizing an advantage mask that is perfectly aligned with the shared prefix.

\subsubsection{Self-sequence Sharing.}
Self-sequence sharing mechanism addresses redundancy of KV caches within a single long trajectory and mitigates pipeline inefficiencies caused by workload imbalances~\cite{wang2025wlb}. Conceptually similar to the mechanism of chunked prefill~\cite{agrawal2023sarathi} in inference systems, this mechanism allows to partition a long sequence into multiple contiguous chunks and schedule them across different micro-batches. As shown in Fig.~\ref{fig:Shared_KV}(b), the prefix chunk is processed in an earlier micro-batch, and the system preserves the corresponding KV states within the cache. When a subsequent chunk is executed in a later micro-batch, the system retrieves the cached KV states to initialize the attention computation, thereby bypassing the recomputation of the prefix.

% \section{Data Scheduling}
\subsection{Workload Balance with Semantic Grouping}
To fully leverage the capability of the developed KV sharing mechanisms, it is essential to balance the workload evenly among distributed workers while simultaneously maximizing the rate of prefix reuse. To achieve this goal, \Sys first constructs semantic groups that cluster training data based on prefix similarity, thereby consolidating sequences with high potential for reuse.

\noindent\textbf{Semantic grouping.}\label{sec:Semantic Grouping}
Structured RL sampling inherently groups sequences by their semantic origin—for example, multiple generation branches from the same prompt or successive samples collected along a shared trajectory. We refer to each such cluster as a \emph{semantic group}. Sequences within the same group are not only semantically related but also tend to exhibit substantial token prefix overlap, since they originate from the same prompt or share a common trajectory prefix. To validate this observation, we sort training sequences by semantic origin and measure pairwise prefix matching rates within a batch. Our analysis confirms a clear separation: high prefix overlap is concentrated within groups, whereas cross-group overlap remains limited. This indicates that prefix reuse is predominantly a within-group phenomenon.

Motivated by this finding, \Sys adopts semantic groups, rather than individual sequences, as the fundamental unit of scheduling. By keeping a group intact during placement, \Sys preserves most prefix-sharing opportunities without incurring the complexity of fine-grained cross-worker coordination. This design also provides a natural abstraction for workload estimation: the scheduler can compute each group’s effective workload by aggregating sequence lengths after accounting for cacheable prefixes, and then strategically place groups across workers to improve both load balance and KV-cache locality.

\noindent\textbf{Load balancing.}
To enable efficient partitioning of semantic groups across training workers, \Sys estimates the computational workload of each group based on its constituent sequences. For a given sequence $s$ with length $l$ and cached prefix length $h$, the workload is determined by the uncached suffix of length $l^{suf} = l - h$. \Sys models the execution time $T(s)$ as a combination of attention computation, MLP processing, and system overhead:
\vspace{-.5em}
\begin{equation}
T(s)=\alpha_1\cdot l^{suf}\cdot l+\alpha_2\cdot l^{suf}+\alpha_3,
\vspace{-.5em}
\label{eq:single_sequence}
\end{equation}
where $\alpha_1$, $\alpha_2$, and $\alpha_3$ are associated parameters derived from profiling. The workload of a semantic group is then computed as the sum of $T(s)$ on all sequences within that group.

With group-level workload estimates, \Sys performs load-balanced partitioning by greedily placing semantic groups across workers. Specifically, semantic groups are sorted in descending order of estimated workload and iteratively assigned to the worker with the smallest cumulative load. After each assignment, the worker's load is updated to reflect the added group cost. Each semantic group is kept intact, i.e., never split across workers, thereby preserving the locality required for high KV-cache reuse.

\begin{figure}[t]
    \centering
    \includegraphics[width=1\linewidth]{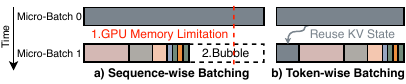}
    \vspace{-1.5em}
    \caption{Comparison of Micro-Batching paradigms. a) Traditional sequence-wise batching treats trajectories as indivisible units. b) Token-wise Micro-Batching partitions long sequences at the token level across micro-batch boundaries.}
    \vspace{-.5em}
    \label{fig:token_level_schedule}
\end{figure}

\begin{figure*}
    \centering
    \includegraphics[width=1\linewidth]{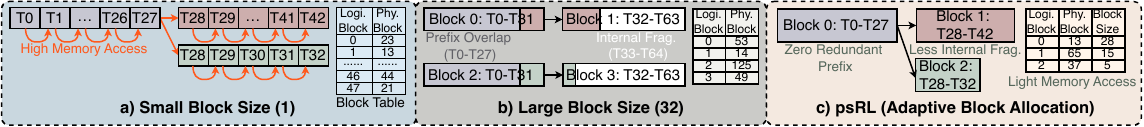}
    \vspace{-1.8em}
    \caption{Comparison of KV block allocation strategies. a) Token-level fine-grained allocation  increases memory access overhead. b) Fixed-granularity allocation with a large block size leads to redundant prefix materialization and internal fragmentation. c) Adaptive block allocation dynamically sizes blocks to exactly cover reusable prefix spans.
    % achieving zero redundant prefixes, less internal fragmentation, and light memory access.
    }
    % \vspace{-.7em}
    \label{fig:Block_Management}
\end{figure*}

\begin{figure}[!h]
\centering\includegraphics[width=1\linewidth]{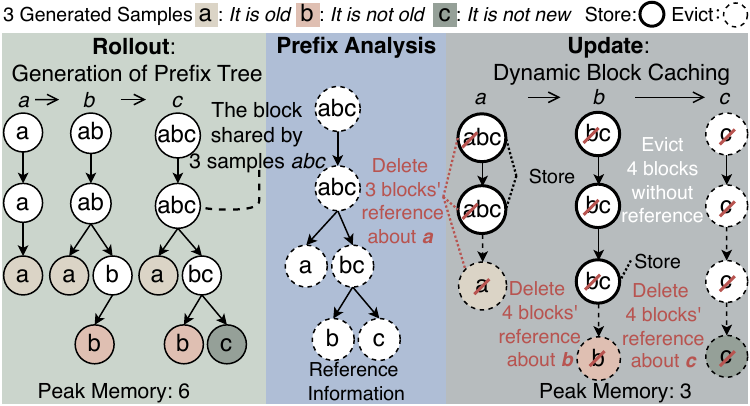}
\vspace{-1.8em}
    \caption{Illustration of Dynamic Block Caching. Left: during rollout, generated samples form a shared prefix tree in which each node corresponds to a KV block and common prefixes are merged. Middle: before training, \Sys analyzes the tree to compute.
    % , defined as the number of future training samples that will access the block. 
    Right: during update, reference counts are decremented at micro-batch granularity as samples are processed. 
    % A block remains in GPU memory while its reference count is positive.
    % and is immediately evicted once the reference count reaches zero. 
}
    \vspace{-1em}
    \label{fig:Generation_Prefix_Tree}
\end{figure}

\subsection{Token-wise Micro-Batching}
With the training workload assigned to each worker, \Sys leverages a fine-grained micro-batching strategy that packs samples into micro-batches to minimize pipeline bubbles. Building on the self-sequence sharing mechanism, this strategy adopts token-wise micro-batching instead of conventional sequence-wise batching. As illustrated in Fig.~\ref{fig:token_level_schedule}, token-wise batching offers distinct advantages under the long-tail workload distributions commonly observed in RL rollouts~\cite{zhong2025streamrl,team2025kimi,he2025history,fu2025areal}. First, it strictly bounds the peak memory footprint of each micro-batch, effectively preventing out-of-memory errors caused by extremely long trajectories. Second, it enables finer-grained workload scheduling by allowing flexible token-level packing across micro-batches.

\Sys partitions the assigned training data into micro-batches based on estimated workload. Given a pre-assigned mini-batch and a given number of micro-batches, \Sys first estimates the total workload of the mini-batch using Eq.~\eqref{eq:single_sequence} and derives a per-micro-batch workload budget. As it processes incoming sequences, it continuously tracks the accumulated workload of the current micro-batch. When adding a new sequence would exceed the remaining budget, \Sys splits the sequence at the token level and distributes its remaining tokens across subsequent micro-batches, thereby ensuring balanced and efficient execution.

\section{Memory Management}
This section presents the details of the memory management mechanisms within \Sys.

\subsection{Adaptive Block Allocation}
As highlighted in \cref{challenges}, existing memory management systems based on fixed-size block allocation are ill-suited for the highly variable lengths of reused prefixes under \Sys. While extremely fine-grained allocation eliminates internal fragmentation, it significantly increases memory access frequency and the overhead of pointer indirection, as illustrated in Fig.~\ref{fig:Block_Management}(a). Conversely, adopting a large block size leads to substantial memory waste. Since prefix sharing operates strictly at block granularity, overlapping prefixes that do not align with block boundaries cannot be effectively reused. For example, as shown in Fig.~\ref{fig:Block_Management}(b), although Block 0 and Block 2 share identical tokens from positions 0 to 27, their divergence at tokens 28 to 31 prevents them from sharing the same block. Consequently, the KV states for the overlapping prefix must be redundantly materialized in both blocks, severely limiting effective prefix reuse and inflating memory consumption.

\Sys is founded on a key insight that enables near-maximal utilization of KV prefixes: in training workloads, the structure of prefix reuse is known in advance, and the training samples themselves remain immutable throughout execution. By analyzing the prefix tree constructed from rollout data, \Sys can precisely identify the span and reuse degree of each shared prefix. This enables block allocation to be guided by semantic reuse boundaries rather than fixed token counts. Leveraging this insight, \Sys introduces Adaptive Block Allocation, which dynamically tailors block sizes to exactly match reusable prefix spans. Each logical block corresponds to a contiguous prefix segment shared across one or more sequences, and its physical allocation is sized according to the actual prefix length. As shown in Fig.~\ref{fig:Block_Management}(c), this design eliminates redundant materialization of prefix states, substantially reduces internal fragmentation, and minimizes unnecessary memory accesses.

\vspace{-.5em}
\subsection{Dynamic Block Caching}\label{sec:Dynamic Block Caching}
\Sys proposes a new KV block caching mechanism driven by prefix tree analysis. As shown in Fig.~\ref{fig:Generation_Prefix_Tree}(left), during rollout, all generated samples jointly form a prefix tree where each node corresponds to a KV block and shared prefixes naturally merge. Before the update phase, \Sys analyzes this tree and, for each KV block, identifies its \emph{reference count}—the number of training samples that will access this block in the subsequent update stage. In Fig.~\ref{fig:Generation_Prefix_Tree}, the reference count of the root block is 4, as all four training samples will access it, while the reference count of the leaf block is 1. This reference count serves as an indicator of each block's reuse potential during training.

During training, \Sys maintains block liveness at micro-batch granularity by decrementing the reference count whenever a sample completes computation on the block. After each micro-batch backward pass, the system removes the corresponding reference entries from all accessed blocks. A block is kept in GPU memory only if its reference count is positive, indicating future reuse.
% ; otherwise, it is immediately evicted. 
In Fig.~\ref{fig:Generation_Prefix_Tree}, when training sample \textit{a}, the first two blocks are preserved because they will be reused by subsequent samples \textit{bcd} and \textit{bc}, while the third block is removed. By caching only blocks that are guaranteed to be reused, \Sys significantly reduces the peak memory footprint without introducing recomputation.

When GPU memory becomes insufficient, \Sys may proactively evict blocks whose reference count is still positive. Because the full prefix tree and the execution order of data expose the exact future access pattern, the system can determine the next access time of each block and evict the one with the farthest reuse distance. This eviction strategy minimizes recomputation and offloading overhead. 
% \vspace{-.5em}
\section{System Implementation}
We implement \Sys on top of veRL~\cite{sheng2025hybridflow}, with the primary system modifications made to the underlying training engine, Megatron-LM~\cite{narayanan2021efficient}. The implementation comprises approximately 7,000 lines of code and is organized around four components: a prefix-aware data orchestrator, a token-wise execution planner, an adaptive KV cache manager, and a prefix-sharing training runtime. Beyond the update phase itself, the same mechanisms are also extended to other RL pipeline stages that operate over the same static training samples and exhibit similar prefix redundancy.
 
\noindent\textbf{Prefix-aware data orchestrator.}
During training initialization, \Sys designates a master device within the update cluster as orchestrator to collect and reorganize rollout samples.  Samples sharing a \texttt{uid} form a semantic group; the master device then partitions and dispatches these groups across devices in different pipelines.

\noindent\textbf{Token-wise execution planner.}
% To minimize computational overhead, 
The planner executes exclusively on the first-stage device of each pipeline. It partitions samples assigned by the orchestrator into micro-batches, analyzing their prefixes to generate the requisite attention, advantage, and loss masks.  These generated masks, alongside the execution sequence, are subsequently dispatched to downstream devices within the same pipeline.

\noindent\textbf{Adaptive KV cache manager.} \Sys implements manager for both block allocation and caching. To reduce memory consumption, KV memory for each variable-sized block is allocated online as a CUDA tensor through the PyTorch caching allocator, rather than carved out from a pre-allocated fixed-size pool. To mitigate allocation overhead, \Sys pre-allocates the required memory one micro-batch ahead of use and overlaps memory creation with ongoing computation. The manager maintains per-block metadata, including the base address, block size, position of the corresponding token segment in the full sequence, and reference count. These metadata allow the runtime to correctly access the blocks, while the reference count supports dynamic block eviction.
% To support attention over variable-sized blocks, \Sys modifies the attention backend to use above metadata to read the block and compute.

\noindent\textbf{Prefix-sharing training runtime.}
For each device, the training runtime executes the ordered micro-batches generated by the planner, and handles the corresponding computation and communication. To support attention over variable-sized blocks, \Sys modifies the attention backend to read the KV-block metadata provided by the memory manager and use them to locate the corresponding KV states for computation.
To avoid redundant backward computation on shared prefix states, we first accumulate all consumer-induced gradient adjoints on the shared prefix representations and then perform a single backward pass through the prefix producer graph.

\noindent\textbf{Accelerating RL pipeline.}
\Sys is also designed to optimize the \emph{reference model forward pass} and \emph{old policy log-probability computation} phases in the RLHF pipeline~\cite{schulman2017proximal,shao2024deepseekmath}. These phases contribute  10\%-20\% of end-to-end training time. They are structurally similar to the update phase because they operate over the same static training samples and require forward computation on highly redundant prefixes. As a result, \Sys extends its prefix sharing, scheduling, and memory management mechanisms to these phases.

\section{Evaluation}

\subsection{Experiment Setup}\label{sec:Experiment Setup}

% \subsubsection{Experiments.}

\subsubsection{Hardware settings.}
We deploy \Sys on a large-scale GPU cluster. Specifically, each node is equipped with four 80\,GB NVIDIA A100 GPUs and powered by the Intel Xeon Gold 6342 CPU with 48 physical cores and 128\,GB of host DRAM. 
Within each node, GPUs are connected via a 200\,Gb/s interconnect, while cross-node communication is supported by a 100\,Gb/s RDMA network fabric. 

\subsubsection{Models and configuration.}\label{sec:Models and configuration}
We evaluate \Sys across a diverse spectrum of agentic RL workloads, ranging from academic benchmarks to large-scale industrial deployments. As summarized in Tab.~\ref{tab:configuration}, our evaluation suite includes the following settings:
1) \textbf{Standard Agents.}
Following the Verl-Agent framework~\cite{feng2025group}, we evaluate agents in the \textit{Search}, \textit{WebShop}, and \textit{ALFWorld} environments. These agents are instantiated using the Qwen2.5~\cite{qwen2025qwen25technicalreport} family with 1.5B and 7B parameters.
2) \textbf{Industrial-Scale Agent.}
To stress-test system scalability, we additionally evaluate a proprietary industrial \textit{DTN Agent}. This agent is a digital twin network operations assistant built on the Qwen3-235B~\cite{yang2025qwen3} Mixture-of-Experts (MoE) model. It is designed as a network engineering assistant that can understand configuration files from routers, firewalls, switches, and other devices from different vendors, and can solve operational tasks using both built-in knowledge and external tools. This agent generates 8 samples for each prompt and supports up to 40K prompt tokens followed by 20K generation tokens per step.

\noindent\textbf{Algorithms.}
We conduct experiment on five RL strategies: Step-wise (\textit{Step})~\cite{feng2025group}, Step with Summarization (\textit{Step+S})~\cite{feng2025group}, Step with Remove Think (\textit{Step+RT)}~\cite{feng2025group}, Tree-structured (\textit{Tree})~\cite{hou2025treerl}, and \textit{Tree-Step}~\cite{ji2025tree}. The setting strictly follows the implementations provided by above frameworks and the most commonly adopted practices for each respective agent.

\noindent\textbf{Distributed execution.}
We align distributed parallelism with model scale to reflect production training environments. For standard benchmarks using Qwen2.5-1.5B and Qwen2.5-7B, we employ a compact 3D parallelism setup with Data Parallelism (DP)=2, Tensor Parallelism (TP)=2, and Pipeline Parallelism (PP)=2, spanning 8 GPUs. For the industrial DTN Agent based on Qwen3-235B, we adopt a larger configuration with DP=4, TP=4, Expert Parallelism (EP)=8, and PP=12, distributing the model across 1536 GPUs.

\begin{table}[t]
    \centering
    \caption{\textbf{Experiment Configurations}}\label{tab:configuration}
        \vspace{-.5em}
    \resizebox{\linewidth}{!}{%
    \begin{tabular}{l l c c c c}
        \toprule
        \textbf{Agent} & 
        \textbf{\makecell[c]{Model}} & 
        \textbf{\makecell[c]{Max \\ Prom.}} & 
        \textbf{\makecell[c]{Max \\ Gen.}} & 
        \textbf{\makecell[c]{Max \\ Step}}  &
        \textbf{\makecell[c]{N \\ Roll. }} \\
        \midrule
        Search       & Qwen2.5-7B   & 4K  & 512    & 4  & 5 \\
        WebShop      & Qwen2.5-1.5B & 4K  & 512    & 15  & 4 \\
        ALFWorld     & Qwen2.5-1.5B & 4K  & 512    & 50  & 8 \\
        DTN Agent & Qwen3-235B   & 40K & 20K & 10 & 8 \\
        \bottomrule
    \end{tabular}
    }
    \vspace{-1em}
\end{table}

\begin{figure*}[ht]
\centering\includegraphics[width=1\linewidth]{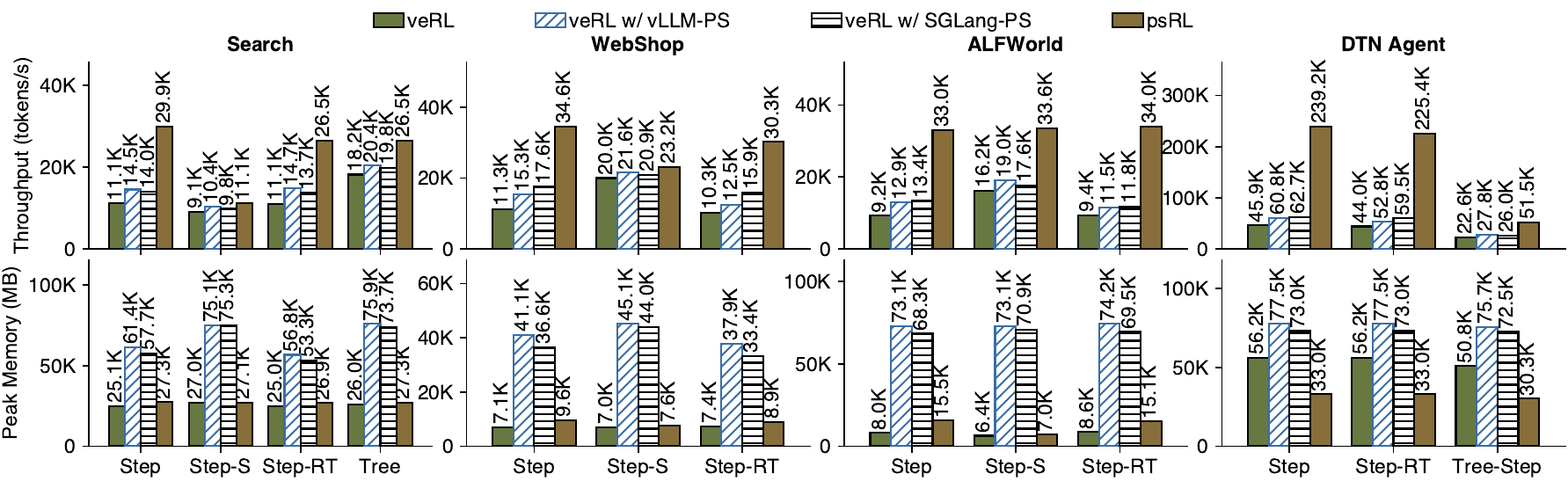}
 \vspace{-2em}
\caption{Performance comparison of the \textit{Update} phase. We report the throughput (tokens/s) and peak GPU memory consumption (MB) across diverse agentic workloads and structured sampling modes.}
    \label{fig:main_results_combined}
    \vspace{-.5em}
\end{figure*}

\subsubsection{Baselines}
We evaluate \Sys against three baselines within the veRL~\cite{sheng2025hybridflow}, whose update phase is powered by Megatron-LM~\cite{narayanan2021efficient}. Because existing training engines do not natively support prefix sharing for these workloads, we compare against the default veRL and two veRL variants whose training engines are augmented with prefix-sharing mechanisms adapted from inference systems~\cite{kwon2023efficient,zheng2024sglang}.

\noindent\textbf{veRL~\cite{sheng2025hybridflow}.}
The default veRL pipeline serves as our primary baseline. Its update phase uses Megatron-LM~\cite{narayanan2021efficient} as the underlying distributed training engine and supports DP, TP, PP, and EP, but without prefix sharing.

\noindent\textbf{veRL w/ vLLM-PS.}
We implement a vLLM-style prefix-sharing mechanism in the update engine of veRL to enable prefix reuse. It adopts fixed-size KV blocks with a block size of 16 tokens, matching the default design of vLLM~\cite{kwon2023efficient}.

\noindent\textbf{veRL w/ SGLang-PS.}
Similarly, we implement an SGLang-style prefix-sharing mechanism in the update engine of veRL. The key difference is  SGLang~\cite{zheng2024sglang} uses a block size of 1 token, which reduces internal fragmentation and improves fine-grained reuse, but incurs higher memory-access overhead.

\subsection{Main Results}
\noindent\textbf{Update throughput.} Fig.~\ref{fig:main_results_combined} depicts the comparison of end-to-end training throughput against veRL, vLLM-PS, and SGLang-PS across four benchmarks.  \Sys consistently dominates, achieving 1.2$\times$--5.2$\times$ speedups over veRL by maximizing prefix reuse. The advantage is most pronounced in the DTN Agent environment under the Step mode, where \Sys reaches 239.1k tokens/s, yielding an approximate 3.8$\times$ speedup over the strongest baseline, SGLang-PS, which peaks at 62.7k tokens/s.  In other agent environments such as Search and WebShop, \Sys generally delivers a 1.5$\times$ to 2.5$\times$ throughput improvement over vLLM-PS and SGLang-PS. This consistent superiority validates the effectiveness of the proposed prefix sharing mechanisms in eliminating redundant computation during the update phase.

\begin{figure}
    \centering
    \includegraphics[width=1\linewidth]{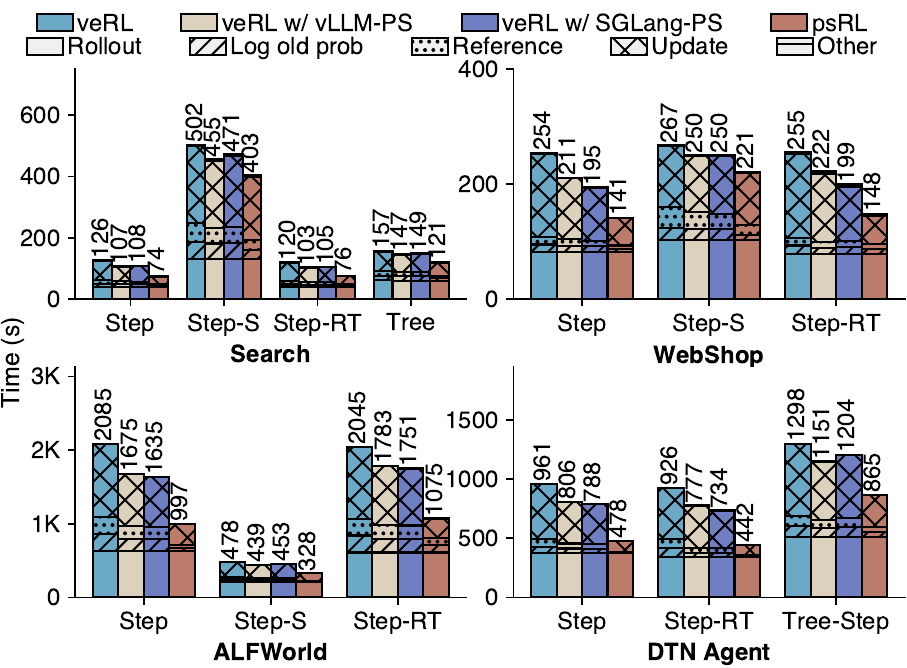}
    \vspace{-2em}
    \caption{E2E Latency comparison of RL training pipeline.}
    \label{fig:E2E}
    \vspace{-1em}
\end{figure}

\noindent\textbf{Memory usage.} Fig.~\ref{fig:main_results_combined} shows peak memory allocations across all systems.  Notably, vLLM-PS and SGLang-PS consistently consume 2$\times$--10$\times$ more memory than veRL; for instance, in ALFWorld (Step-S), vLLM-PS peaks at 73.1\,GB versus veRL's 6.4\,GB. Their inference-centric allocation lacks optimizations vital for training. In contrast, \Sys maintains a minimal memory footprint comparable to veRL across standard benchmarks. Specifically, on the extreme-scale DTN Agent workload, \Sys reduces peak memory from 56.2\,GB to 33\,GB. This 41\% reduction proves our block management effectively eliminates redundant storage in long-context RL, freeing memory capacity for larger batch sizes.

\noindent\textbf{End-to-End RL performance.} Fig.~\ref{fig:E2E} details the execution time breakdown for a complete RL training iteration under veRL~\cite{sheng2025hybridflow}. Overall, \Sys achieves up to 2.1$\times$ end-to-end speedup over veRL, eliminating inherent RLHF computational bottlenecks. In DTN Agent (Step), \Sys cuts Log Old Prob and Reference computation times from 57.1\,s and 68.3\,s to 7.1\,s and 8.3\,s. This 8$\times$ speedup transforms these compute-heavy steps into lightweight, memory-bound operations. Similarly, in ALFWorld, the latency of these two phases drops by over 5$\times$. Furthermore, token-wise micro-batching and dynamic block caching dramatically reduce update time; for DTN Agent, update latency drops from 462.1\,s (veRL) to 88.7\,s. While vLLM-PS and SGLang-PS offer marginal improvements over veRL via basic KV caching, they consistently underperform \Sys. For instance, \Sys completes a WebShop (Step) iteration in just 148.1\,s, significantly outpacing all baselines.

\begin{figure}[t]
    \centering
    % --- (a) Reward Comparison ---
    \begin{subfigure}{0.235\textwidth}
        \centering
        \includegraphics[width=\linewidth]{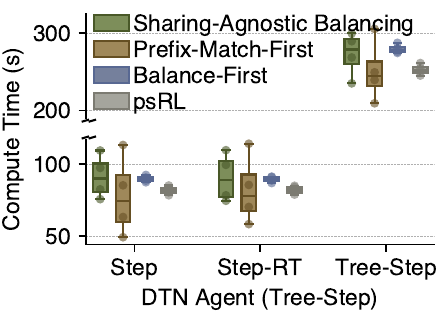}
        \caption{Impact of workload scheduling on the execution time.}
\label{fig:ablation_Workload_Schedule}
    \end{subfigure}
    \begin{subfigure}{0.235\textwidth}
        \centering
        \includegraphics[width=\linewidth]{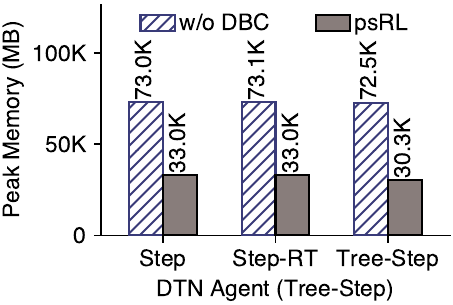}
        \caption{Impact of Dynamic Block Caching (DBC) on peak memory.}
        \label{fig:ablation_DBR}
    \end{subfigure}\vspace{-.5em}
    \caption{Ablation study on individual design of the \Sys.}\label{fig:Ablation}
    % \vspace{-.5em}
\end{figure}

\subsection{Evaluating Individual Design}
To understand the benefits of \Sys, we analyze its design through both component-wise ablation and fine-grained runtime traces. As illustrated in Fig.~\ref{fig:Ablation} and Fig.~\ref{fig:Combined_Depth_Analysis}, we examine how the key designs of \Sys enhance runtime behavior across two complementary dimensions: workload scheduling and memory management.

\subsubsection{Workload scheduling}
The design of workload scheduling within \Sys consists of two components: workload balance across workers and Token-wise Micro-Batching (TMB). Together, they make the execution time of different workers more balanced while also reducing the latency of each worker, thereby improving overall training efficiency.

\noindent\textbf{Workload balance across workers.}
Fig.~\ref{fig:ablation_Workload_Schedule} compares four data scheduling strategies by execution time of each DP worker under DTN Agent. Sharing-Agnostic Balancing, the default strategy of veRL, balances only by sequence length and ignores prefix reuse, leading to substantial execution skew, with worker times in Tree-Step ranging from 235.12\,s to 299.45\,s. Prefix-Match-First maximizes prefix reuse but introduces even more severe imbalance; in Tree-Step, worker times range from 209.45\,s to 305.12\,s. Balance-First always assigns data to the currently lightest worker, which improves balance but sacrifices prefix locality, resulting in a maximum worker time of 287.25\,s in Tree-Step. In contrast, \Sys jointly optimize prefix reuse and workload balance, with worker times tightly grouped between 245.13\,s and 261.23\,s in Tree-Step.

\begin{figure}[t]
    \centering
\includegraphics[width=1\linewidth]{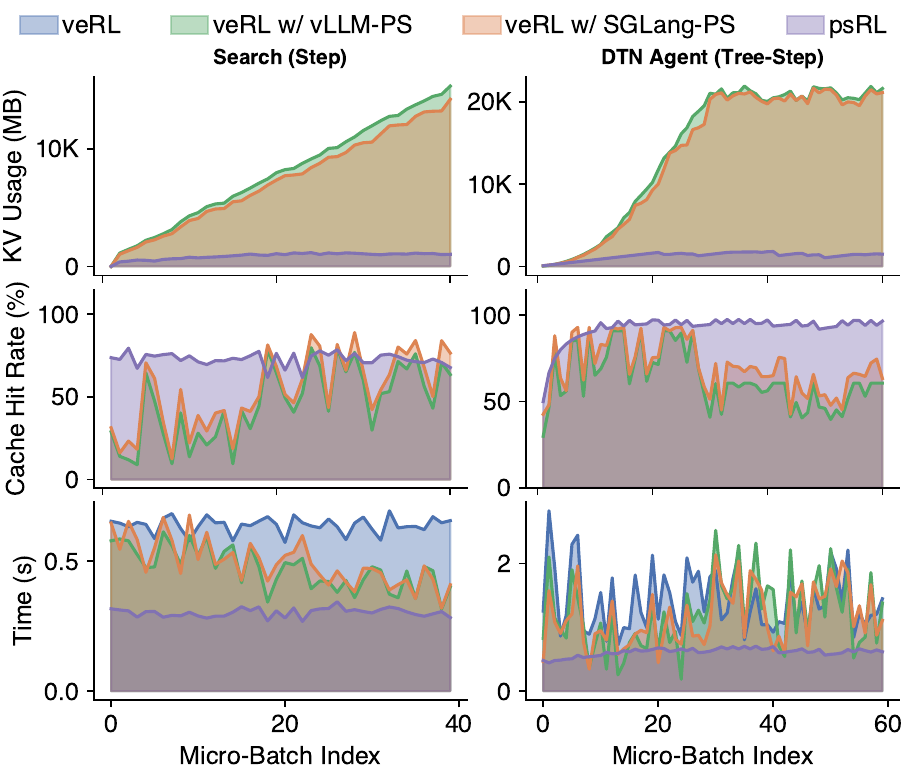}
    \vspace{-2em}
    \caption{ Micro-batch level dynamics of KV memory allocation, cache hit rate and execution latency.}
    \vspace{-1em}
\label{fig:Combined_Depth_Analysis}
\end{figure}

\noindent\textbf{Token-wise micro-batching.}
At a finer granularity, TMB further smooths execution within each worker by repacking and partitioning data according to token-level computational density. As shown in Fig.~\ref{fig:Combined_Depth_Analysis} (bottom), this design makes the forward and backward latency of each micro-batch much more uniform, avoiding severe latency spikes and jitter. Such smoothing is critical for distributed training, because it reduces wait-time bubbles and keeps GPU compute units consistently busy throughout the training iteration. As a result, TMB not only shortens per-worker execution time, but also improves global throughput by making micro-batch execution more predictable and balanced.

\subsubsection{Memory management}
The memory management of \Sys consists of Adaptive Block Allocation (ABA) and Dynamic Block Caching (DBC). ABA improves memory efficiency and access efficiency at the allocation level, while DBC controls KV-state growth and preserves high cache effectiveness during execution.

\noindent\textbf{Dynamic block caching.}
DBC further improves memory behavior during execution. As shown in Fig.~\ref{fig:ablation_DBR}, without DBC, KV-state memory grows excessively in DTN Agent, reaching 73.0k\,MB and 73.1k\,MB in Step and Step-RT modes. With DBC, peak memory consumption drops to 33.0k\,MB in both modes, a reduction of more than 54\%. In Tree-Step mode, memory usage similarly falls from 72.5k\,MB to 30.3k\,MB. This reduction comes from reclaiming blocks immediately after their final use rather than retaining redundant states. 

The fine-grained runtime trace in Fig.~\ref{fig:Combined_Depth_Analysis} (top and middle) further shows that DBC controls KV-state growth without reducing cache effectiveness. Traditional prefix-sharing mechanisms such as vLLM-PS and SGLang-PS lack fine-grained lifecycle semantics for extreme-context RL trajectories. In DTN Agent (Tree-Step), their KV-state footprint grows nearly linearly during the initial phase, then stalls and fluctuates sharply around the 30th micro-batch due to OOM pressure and reactive mechanisms such as forced cache eviction. In contrast, \Sys uses DBC to reclaim invalidated KV blocks after their final referencing computation step, producing a controlled sawtooth KV-memory pattern that bounds peak usage within hardware limits. More importantly, this bounded footprint does not compromise cache utilization.

\begin{figure}
    \centering
    \includegraphics[width=\linewidth]{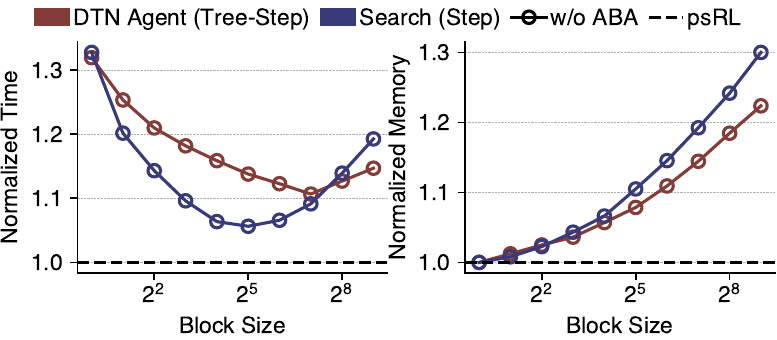}
    \vspace{-2em}
    \caption{Performance comparison between Adaptive Block Allocation (ABA) and fixed-size block strategies, normalized to \Sys.}
\label{fig:ablation_blocksize}\vspace{-.5em}

\end{figure}

\noindent\textbf{Adaptive block allocation.}
As shown in Fig.~\ref{fig:ablation_blocksize}, fixed-size paging exposes a fundamental tradeoff between memory efficiency and execution speed, and the optimal block size varies significantly across workloads. As block size increases from 1 to 512, peak memory consumption rises monotonically because of internal fragmentation, increasing from 27k\,MB to 35k\,MB in the Search environment and reaching a 1.3$\times$ normalized memory overhead. Execution time, however, follows a U-shaped trend: the best fixed block size is 32 in Search but 128 in DTN Agent. This sensitivity makes manual tuning impractical. In contrast, ABA allocates blocks according to reusable prefix spans, which reduces internal fragmentation, improves memory access efficiency, and increases effective prefix reuse without requiring workload-specific tuning.

\subsection{System Scalability}
\begin{figure}[t]
    \centering
    \includegraphics[width=1\linewidth]{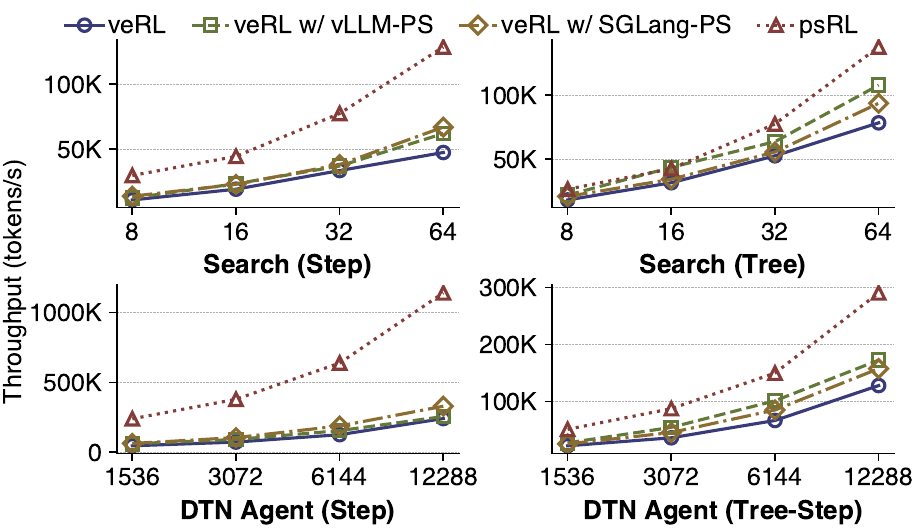}
    \vspace{-2em}
    \caption{Throughput of \Sys against baseline systems across varying cluster sizes.}
    % \vspace{-1em}
    \label{fig:ablation_device_num}
\end{figure}

\noindent\textbf{Scalability w.r.t. \#devices.}
We evaluate \Sys's scalability by proportionally doubling the DP degree alongside the GPU count. Fig.~\ref{fig:ablation_device_num} demonstrates near-linear scaling across both small and extreme cluster setups. On the Search benchmark, \Sys maintains a clear lead. In Step mode on 64 devices, it achieves 128.5k tokens/s, outperforming veRL by 2.7$\times$ and SGLang-PS by 1.9$\times$. Under extreme-scale conditions testing the DTN Agent, \Sys exhibits highly robust scalability. On 12,288 GPUs, our throughput reaches 1.14M tokens/s, dwarfing veRL by 4.7$\times$ and SGLang-PS by 3.4$\times$. These results confirm that our memory and scheduling designs introduce negligible overhead, unlocking the full computational potential of massive distributed clusters.

\begin{figure}[!t]
    \centering
\includegraphics[width=1\linewidth]{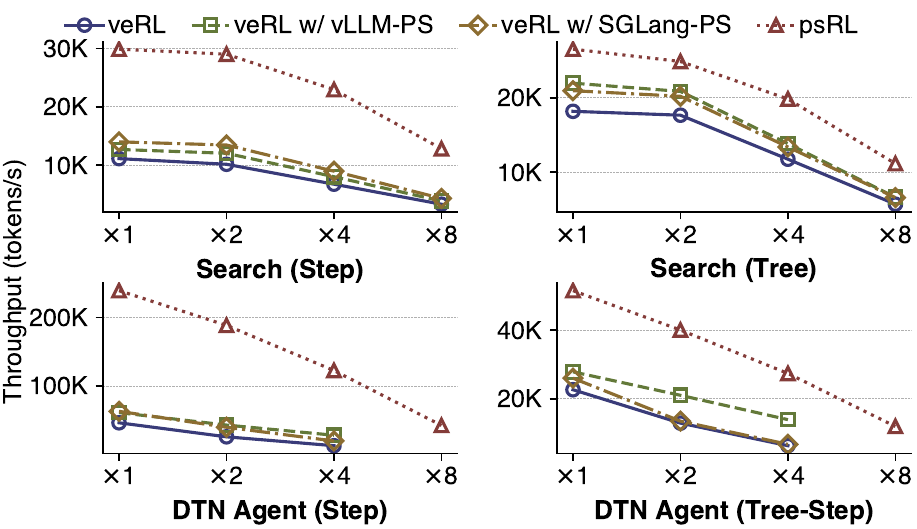}
\vspace{-2em}
    \caption{Throughput of \Sys against baseline systems across varying scales of sequence lengths.}
    % \vspace{-1.em}
\label{fig:ablation_seqlen}
\end{figure}

\noindent\textbf{Scalability w.r.t. sequence length.}
We evaluate system resilience to longer sequences by scaling sample length from 1$\times$ to 8$\times$. To isolate this effect without altering prefix reuse rates, we insert $n$-1 dummy tokens per original token for $n\times$ scales. Fig.~\ref{fig:ablation_seqlen} shows the resulting throughput. As attention computation grows quadratically, \Sys degrades gracefully despite steep baseline drop-offs. In the Search benchmark (Step) at 8$\times$, \Sys sustains 12.8k tokens/s, maintaining a 3$\times$--4$\times$ advantage over SGLang-PS and veRL. Crucially, on DTN Agent at 8$\times$ length, immense memory pressure from long-context causes all baselines to exhaust GPU memory and crash. In contrast, \Sys survives, delivering 42.1k tokens/s in Step mode.

% \vspace{-.3em}
\section{Related Works}
\noindent\textbf{Agentic systems.} 
Early systems such as NeMo-Aligner~\cite{shen2024nemo} and OpenRLHF~\cite{hu2024openrlhf} adopt a disaggregated pipeline architecture, where rollout and training stages are executed on separate resources. Frameworks such as veRL~\cite{sheng2025hybridflow}, ReaL~\cite{mei2024realhf}, and RLHFuse~\cite{zhong2025optimizing} colocate multiple RL stages within the same GPU pool to better overlap computation and reduce idle time.  StreamRL~\cite{zhong2025streamrl} proposes a one-step off-policy pipeline where rollout proceeds with slightly stale policy weights, enabling better overlap between generation and training. AReaL~\cite{fu2025areal} further relaxes synchronization constraints and enables fully asynchronous RL training. Building on this paradigm, RhymeRL~\cite{he2025history} integrates speculative decoding to accelerate rollout generation by leveraging history output. 
Recent work like
Tree Training~\cite{wang2026treetrainingacceleratingagentic} and
AReaL-DTA~\cite{zhang2026arealdtadynamictreeattention} organize samples as
  explicit prefix trees and reuse computation through DFS-based execution. In contrast, \Sys supports intra-batch, inter-batch, and self-sequence sharing, decoupling prefix reuse from tree execution and jointly optimizing scheduling
  and memory management.

\noindent\textbf{KV management.} 
A broad line of research heavily optimizes KV cache management for online LLM serving. To mitigate memory fragmentation, vLLM~\cite{kwon2023efficient} introduces PagedAttention, whereas SGLang~\cite{zheng2024sglang} proposes RadixAttention to dynamically eliminate redundant prefix computations.  Subsequent efforts extend these concepts to complex serving scenarios, including prefix caching for chatbots~\cite{gao2024cost}, streaming-aware scheduling~\cite{yu2025stateful}, and cache management for tool calling (e.g., InferCept~\cite{abhyankar2024infercept}). Furthermore, systems such as Autellix~\cite{luo2025autellix} and ParrotServe~\cite{lin2024parrot} specifically explore request scheduling for agentic workflows during inference. Furthermore, Jenga~\cite{zhang2025jenga} employs an attention-aware allocator and pattern-based eviction strategies to enhance memory reuse, while DiffKV~\cite{zhang2025diffkv} introduces an on-GPU memory manager that compacts fragmented memory in parallel, translating cache sparsity into performance gains.

\section{Conclusion}

This paper presents \Sys, a prefix-sharing training system designed for modern agentic RL workloads. We observe that the bottleneck in RL training is shifting from rollout to update, as tree-structured and step-wise RL strategies substantially increase training sample volume with only a low marginal cost for rollout. Moreover, we find that these sampling strategies introduce significant prefix redundancy among training samples—an opportunity that existing training systems largely fail to exploit. \Sys addresses this gap by leveraging global visibility and data immutability to enable prefix-aware optimizations in both workload scheduling and memory management. Specifically, \Sys introduces novel mechanisms for flexible prefix sharing, load-balanced execution, adaptive KV allocation, and dynamic KV caching. Extensive evaluations using production traces and real-world agentic workloads demonstrate that \Sys improves agentic training throughput by up to 5.2$\times$.

\bibliographystyle{acm_reference_format}
\bibliography{references}
\clearpage

\end{document}